\documentclass[11pt]{article}

\usepackage{amssymb,amsfonts,amsmath,nicefrac,lscape}
\usepackage{bm}
\usepackage[dvips]{epsfig}

\newcommand{\es}{\mbox{\bf{E}}}

\newcommand{\defeq}{\stackrel{\mathrm{.}}{=}}

\newcommand{\pab}{\textsc{ptcb}}

\newcommand{\pabdef}{Price-Transition Cost Balanced}

\newcommand{\udc}{\textsc{utc}}

\newcommand{\udcdef}{Unit Transition Cost}

\newcommand{\bp}{\textsc{bsi}}

\newcommand{\bpdef}{Bregman-Slutsky Inputs}

\newcommand{\epf}{\textsc{epf}}
\newcommand{\epfn}{EPF}

\newcommand{\lda}{\textsc{lda}}
\newcommand{\ldan}{LDA}
\newcommand{\ldadef}{Lowest Distortion Aggregator}

\newtheorem{thm}{Theorem}
\newtheorem{lem}{Lemma}

\newtheorem{defn}{Definition}

\def\cqfd{\hfill\hbox{$\hbox{\vrule width 0.8pt
\vbox to6pt{\hrule depth 0.8pt width 5.2pt
\vfill\hrule depth 0.8pt}\vrule width 0.8pt}$}}

\title{Information geometries for microeconomic theories}

\author{Richard Nock\\
$^*$Centre d'Etude et de Recherche en Economie, Gestion, Mod\'elisation\\
et Informatique Appliqu\'ee (\textsc{Ceregmia} --- UAG),\\
PO Box 7209, Schoelcher 97275, France.\\
\texttt{rnock@martinique.univ-ag.fr}
\and
Brice Magdalou$^*$\\
\texttt{brice.magdalou@martinique.univ-ag.fr}
\and
Nicolas Sanz\\
\textsc{Ceregmia} --- UAG,\\
PO Box 792, Cayenne 97400, France.\\
\texttt{Fred.Celimene@martinique.univ-ag.fr}
\and
Eric Briys$^*$\\
\texttt{eric.b@cyberlibris.com}
\and
Fred C{\'e}lim{\`e}ne$^*$\\
\texttt{Fred.Celimene@martinique.univ-ag.fr}
\and
Frank Nielsen\\
LIX --- Ecole Polytechnique, Palaiseau 91128, France\\
$\&$ Sony Computer Science Laboratories Inc., 3-14-13\\
Higashi Gotanda, Shinagawa-Ku, 141-0022 Tokyo, Japan.\\
\texttt{Nielsen@acm.org}}

\begin{document}


\maketitle

\begin{abstract}
More than thirty years ago, Charnes, Cooper and Schinnar (1976) established
an enlightening contact between economic production functions (\epf s) --- a cornerstone
of neoclassical economics --- and information theory,
showing how a generalization of the Cobb-Douglas production function
encodes homogeneous functions.

As expected by Charnes \textit{et al.}, the contact turns out to be much broader: 
we show how information geometry as pioneered by Amari and others underpins
static and dynamic descriptions of microeconomic cornerstones.

We show that the most popular \epf s are fundamentally grounded in
a very weak axiomatization of economic transition costs between inputs. The strength of this characterization is surprising, as
it geometrically bonds altogether a wealth of collateral economic notions
 --- advocating for applications in various economic fields ---: among all, 
it characterizes (i) Marshallian and Hicksian demands and their geometric duality, (ii) Slutsky-type properties
for the transformation paths, (iii) Roy-type properties for their elementary variations.
\end{abstract}

\section{Introduction}
Microeconomic theory builds from the behavior of individual agents --- consumers and producers ---
to compute aggregate economic outcomes \cite{mcwgMT}. A cornerstone of the neoclassical school of
economics consists in performing such aggregations using \textit{Economic Production Functions} (\epf s).
Obliterating the technical underpinnings \cite{sTO}, an \epf~$\mu_x$ simply aggregates a set of inputs $x_1, x_2, ..., x_m$:
\begin{eqnarray}
\mu_x & \defeq & f\left(x_1, x_2, ..., x_m\right)\:\:, \label{defEPF}
\end{eqnarray}
where $\mu_x$ is the output. We use the term
``production'' because of historical reasons \cite{mAB,sTO}, but \epf s can be used to
aggregate any economically relevant inputs, such as consumptions, prices, productions, labors, 
capitals, incomes, etc. \cite{bkMC,dsMC,gmrSF}. As such, \epf s are not only used
to model consumers' and producers' behaviors: they are virtually used in any
branch of economic analysis \cite{sND},
and even other fields as well \cite{cMM,dpgcET}. 

A significant part of microeconomic theory, grounded in the use of these \epf s,
has been widely criticized for its \textit{ad hoc} status, despite longstanding
celebrated successes \cite{sTC}. Accordingly,
\epf s look like solutions in search of a problem. Our intention is not to settle
the debate from a restrictive economics standpoint. It is rather to build upon
an enlightening rationale \cite{ccsAT}, and follow further their information theoretic
Ariadne's thread to the core of microeconomic theories.

We start with a weak axiomatization of \textit{economic transition costs} between
different combinations of inputs, inspired by more recent works in information geometry \cite{anMO,bgwOT}.
 It has dramatic consequences: it grounds the most popular \epf s
as optimal and exhaustive for transition costs from general standpoints; it provides an
exhaustive description of \textit{economic transformation paths} and the algorithmics of
transition. The dynamics of transition surprisingly reproduce properties
well-known for \epf s (\textit{e.g.} \textit{Slutsky}'s and \textit{Roy}'s identities),
and patch the expansion path to new paths that we call \textit{Hicksian} and \textit{Marshallian}.

Section 2 presents preliminary definition and properties; section 3 gives our main results; a last
section discusses and concludes.
In order not to laden the paper's body, proofs are postponed to an appendix at the end of the paper.

\section{Preliminary definitions and properties}

\subsection{Bregman divergences and \ldan s}

Bold notations such as $\bm{x}$ denote vector-based notations, and blackboard faces such as ${\mathbb{X}}$ sets of (tuples of) real numbers of ${\mathbb{R}}$ or natural integers of ${\mathbb{N}}$. The information-theoretic part of this paper relies on two principal tools: Bregman divergences \cite{bTR}, and Lowest Distortion Aggregators.
\begin{defn}\label{defbreg}
Let $\varphi : {\mathbb{X}} \rightarrow {\mathbb{R}}$ be strictly convex, differentiable over the interior of ${\mathbb{X}}$, with ${\mathbb{X}} \subseteq {\mathbb{R}}^d$ convex:
\begin{itemize}
\item the \textsl{Bregman Divergence} $D_\varphi$ with generator $\varphi$ is \cite{bTR,bgwOT}:
\begin{eqnarray}
D_\varphi(\bm{x}_i || \bm{x}_j)  & \defeq & \varphi(\bm{x}_i) - \varphi(\bm{x}_j) - (\bm{x}_i - \bm{x}_j)^\top \bm{\nabla}_\varphi (\bm{x}_j) \:\:, \label{bdir}
\end{eqnarray}
where $\bm{\nabla}_\varphi(\bm{x}_j) \defeq \left[ \partial \varphi(\bm{x}_j)/\partial x_{jk} \right]^\top$ is the gradient operator;
\item let ${\mathbb{S}} \defeq \{(\bm{x}_i, \gamma_i)\}_{i=1}^{m}$, with $\gamma_i \in {\mathbb{R}_{+*}}, i=1, 2, ..., m$. The \textsl{\ldadef} (\lda) $\bm{\mu}_\varphi$ with generator $\varphi$ for set ${\mathbb{S}}$ is:
\begin{eqnarray}
\bm{\mu}_{\varphi} & \defeq & \bm{\nabla}^{-1}_{\varphi} \left(\frac{1}{\Gamma} \sum_{i=1}^{m} {\gamma_i \bm{\nabla}_\varphi(\bm{x}_i)}\right) \:\:, \Gamma \defeq \sum_{i=1}^{m} {\gamma_i}\:\:.\label{bregmeangen}
\end{eqnarray}
\end{itemize}
\end{defn}
The shape of $\bm{\mu}_{\varphi}$ is that of an Economic Production Function (\epf), in which each $\bm{x}_i$ is a possibly multidimensional input. This is a setting 
far more general than mainstream economics where each $x_i$ would be a scalar input (\ref{defEPF}), ${\mathbb{X}}$ an interval of ${\mathbb{R}}$ ($d=1$), and so $\nabla_\varphi$ the conventional 
derivative. To distinguish this conventional case when $d=1$, we shall use the shorthands for derivatives:
\begin{eqnarray}
\varphi^{[k]}(x) & \defeq & \mathrm{d}^k \varphi(x) / \mathrm{d} x^k, \forall k \in {\mathbb{N}}_* \:\:,\label{dergen}\\
{\textsc{u}}(x) & \defeq & \varphi^{[1]}(x) \:\:. \label{deru}
\end{eqnarray}
The use of (\ref{deru}) instead of the general (\ref{dergen}) is intended to help the reader grasp potential \lda~applications
with utility functions: in economics of risk, $\mu_{\varphi}$ would be the certainty equivalent for
function $\textsc{u}$ \cite{mcwgMT}; in normative economics, $\mu_{\varphi}$ would be the equally distributed
equivalent income for function $\textsc{u}$ \cite{bbdII}, and so on. Notice that multidimensionality is economically interesting, because in this setting each $\bm{x}_i$ could represent a set of capital, a set of labour, etc. . 
However, for the sake of clarity and to remain stick to mainstream, most of the remaining of this paper, starting from the next subsection, is devoted to 
the 1-dimensional input case. Table \ref{t-exe} displays popular Bregman divergences tailored to this setting. 

\epf s are grounded in baseline mathematical properties, such as concavity or convexity that play important roles in shaping preference or aversion 
for diversity \cite{bkMC,dsMC}. Following are baseline properties for \lda s that could be of use in an \epf~setting. A part of the following Lemma is proofsketched in the Appendix.
\begin{lem}\label{lem1}
The following properties hold true for any \lda~$\bm{\mu}_\varphi$:
\begin{itemize}
\item (min - max bounds) $\bm{x}_{\mathrm{min}} \leq \bm{\mu}_\varphi \leq \bm{x}_{\mathrm{max}}$, where $x_{\mathrm{min}, j} = \min_i x_{ij}$ and $x_{\mathrm{max}, j} = \max_i x_{ij}, \forall j = 1, 2, ..., m$;
\item (stability under composition) the composition of \lda s with the same generator $\varphi$ is a \lda~with generator $\varphi$;
\item (invariance modulo linear transforms) let $\varphi_{\bm{b},c}(\bm{x}_i) \defeq \varphi(\bm{x}_i) + \bm{b}^\top \bm{x}_i + c$, with $\bm{b} \in {\mathbb{R}}^d, c\in {\mathbb{R}}$. Then $D_{\varphi_{\bm{b}, c}}(\bm{x}_i || \bm{x}_j) = D_\varphi(\bm{x}_i || \bm{x}_j)$ and $\bm{\mu}_{\varphi_{\bm{b}, c}} = \bm{\mu}_{\varphi}$;
\item (concavity - convexity duality) $\bm{\mu}_\varphi$ is concave if and only if its \textsl{dual} $\bm{\mu}_{\varphi^\star}$ is convex, where $\varphi^\star$ is the Legendre conjugate of $\varphi$:
\begin{eqnarray}
\varphi^\star(\bm{x}_i) & \defeq & \sup_{\bm{y} \in {\mathbb{X}}} \{ \bm{y}^\top \bm{x}_i - \varphi(\bm{y})\} \:\:. \label{deflegc}
\end{eqnarray}
\item (relationship with arithmetic \lda) if $\bm{\mu}_{\varphi}$ is concave (resp. convex), then it is upperbounded (resp. lowerbounded) by the arithmetic \lda: 
\begin{eqnarray}
\bm{\mu} & \defeq & \frac{1}{\Gamma} \sum_{i=1}^{m} {\gamma_i \bm{x}_i}\:\:. \label{defarith}
\end{eqnarray}
\end{itemize}
\end{lem}

\subsection{\ldan s and \epfn s}

We let $p$ be the output price, and $p_i, i=1, 2, ..., m$ the price of input $i$, with $P\defeq \sum_{i=1}^{m} {p_i}$. 

\subsubsection{Marshallian and Hicksian demands on \ldan s}
The following Lemma (proof straightforward) gives an important result in connection with \epf s.
\begin{lem}\label{lem2} The expansion path of concave \lda~$\mu_\varphi$, \textit{i.e.} the solution of $\max_{\{x_i\}_{i=1}^{m}} \{p \mu_\varphi - \sum_{i=1}^{m}{p_i x_i}\}$, is given by the sets of $m$-tuples $\{x_i\}_{i=1}^{m}$ such that:
\begin{eqnarray}
\varphi^{[2]}(x_i) & = & \frac{(p_i/\gamma_i)}{(p_j/\gamma_j)} \varphi^{[2]}(x_j)\:\:, \forall i,j = 1, 2, ..., m\:\:. \label{expath}
\end{eqnarray}
The equivalent problem can be formulated for convex \lda s, after flipping min/max.
\end{lem}
This Lemma admits interesting economic consequences, some beyond the scope of this paper. Here is an example.
\begin{lem}
On the expansion path of any \lda~$\mu_\varphi$, the marginal rate of substitution of $x_i$ for $x_j$ (\ref{defmarsub}) satisfies ${\mathrm{\textsc{s}}}_{\mu_\varphi}^{x_i \rightarrow x_j} = (p_i/\gamma_i) / (p_j/\gamma_j)$, for any $i, j = 1, 2, ..., m$.
\end{lem}
There are two important economic problems related to the optimization of \epf s: maximizing income under a budget constraint, and minimizing expenditures under an output constraint.
Let us cast them for general \lda s.
\begin{defn}\label{defwh}
The Marshallian demand for concave \lda~$\mu_\varphi$ is the problem:
\begin{eqnarray}
\{x_i\}_{i=1}^{m} & = & \arg\max_{\{y_i\}_{i=1}^{m}} {p \textsc{u}^{-1} \left(\frac{1}{\Gamma} \sum_{i=1}^{m} {\gamma_i \textsc{u}(y_i)}\right)} \label{defwalras}\\
 & & \mbox{s.t. } \sum_{i=1}^{m} {p_i y_i} \leq w \nonumber \:\:,
\end{eqnarray}
The Hicksian demand for concave \lda~$\mu_\varphi$ is the problem:
\begin{eqnarray}
\{x_i\}_{i=1}^{m} & = & \arg\min_{\{y_i\}_{i=1}^{m}} \sum_{i=1}^{m} {p_i y_i} \label{defhicks}\\
 & & \mbox{s.t. } p\textsc{u}^{-1} \left(\frac{1}{\Gamma} \sum_{i=1}^{m} {\gamma_i \textsc{u}(y_i)}\right) \geq p\mu' \nonumber \:\:.
\end{eqnarray}
Above, $w>0$ is an income and $\mu'\geq 0$ an output level. Equivalent problems may be formulated for convex \lda s, after flipping min/max and the inequalities.
\end{defn}
The proof of the following Lemma easily follows from Lemma \ref{lem2}. It is an important economic sanity check on \lda s.
\begin{lem}\label{lem3}
For any concave (convex) \lda, the Marshallian and Hicksian demands belong to its expansion path.
\end{lem}

\subsubsection{\pabdef~setting}

Lemma \ref{lem2} says that the expansion path of some \lda~is not necessarily linear in general. Yet, popular \epf s have linear expansion paths, such as Cobb-Douglas. Most remarkably, an important setting produces a linear subspace in the expansion path of \textit{any} \lda. If one makes the assumption that $p_i / \gamma_i$ is a constant for any $i=1, 2, ..., m$ --- a situation to which we refer as \textit{\pabdef} (\pab~for short, rationale in the following Section) ---, then the ${\mathbb{R}}^m$ linear subspace ${\mathbb{L}}$ defined by $x_1 = x_2 = ... = x_m$ inside the domain of $\mu_\varphi$ belongs to the expansion path of $\mu_\varphi$. This is stated below.
\begin{lem}\label{lwh}
The Marshallian and Hicksian demands in the \pab~setting are:
\begin{eqnarray}
x_j & = & \mu \:\:, \forall j=1, 2, ..., m\:\:\mbox{ (Marshallian) }, \label{pabwd}\\
x_j & = & \mu_\varphi \:\:, \forall j=1, 2, ..., m\:\:\mbox{ (Hicksian) }, \label{pabhd}
\end{eqnarray}
where $\mu$ is given in (\ref{defarith}) and $\mu_\varphi$ is given in (\ref{bregmeangen}).
\end{lem}

\section{Main results}

\subsection{Mainstream \epfn s are \ldan s}

A natural question on \lda s is whether they can accurately represent a significant part of mainstream \epf s. We answer affirmatively this question
on the basis of the six categories of \epf s presented on the left column of Table \ref{t-epf} (Theorem also summarized in the Table).
\begin{thm}\label{thop}
The following holds true:
\begin{itemize}
\item [\bf{(I)}] the following \epf s (Table \ref{t-epf}) are \lda s (\ref{bregmeangen}):\\
$\bullet$ Constant Elasticity of Substitution (CES), for generator:
\begin{eqnarray}
\varphi_{\mathrm{\textsc{ces}}}(x) & \defeq & a x^{2 - \frac{1}{\sigma}} + bx + c\:\:; \label{propnces}
\end{eqnarray}
$\bullet$ Cobb-Douglas with constant returns to scale (CD, $\sum_{i} {\beta_i} = 1$), for generator:
\begin{eqnarray}
\varphi_{\mathrm{\textsc{cd}}}(x) & \defeq & a' x\log x + b x + c\label{cobb}\:\:;
\end{eqnarray}
$\bullet$ Generalized Exponential Mean (GEM), for generator:
\begin{eqnarray}
\varphi_{\mathrm{\textsc{gem}}}(x) & \defeq & a' \exp(\theta x + d) + b x + c\label{gengem}\:\:.
\end{eqnarray} 
\item [\bf{(II)}] Leontief is a limit case of \lda, for $\sigma \rightarrow 0^+$ in (\ref{propnces});
\item [\bf{(III)}] Translog and MSTs are not \lda s.
\end{itemize}
Above, $a\in {\mathbb{R}}_*$ is such that (\ref{propnces}) is convex, $a' \in {\mathbb{R}}_{+*}, b, c, d \in {\mathbb{R}}$.
\end{thm}
(proofsketch: see the Appendix). Thus, the most popular \epf s appearing in the theories of the consumer and producer, but also
in normative economics, are in fact \lda \cite{mcwgMT,bbdII}.
CES is the most favourably positioned: with the exception of Arimoto divergences, CES
spans the \lda s corresponding to all Bregman divergences in Table \ref{t-exe}. 

\subsection{Mainstream \epfn s are economically exhaustive for \ldan s}

In this section, we switch to the main analytical economic assumptions that can be made about \epf s, and check which \lda s satisfy them. The (non-empty) subset of \lda s obtained is called ``exhaustive'' for the assumption. To distinguish between different sets of inputs, a general \lda/\epf~for inputs $x_1, x_2, ..., x_m$ shall be denoted $\mu_x$ ($\varphi$ is implicit) (\ref{defEPF}). We summarize these assumptions. The first defines \textit{dually coupled} \epf s $\mu_x$ and $\mu_z$, that satisfy:
\begin{eqnarray}
\sum_{i=1}^{m} {x_i z_i} & = & \mu_x \mu_z \:\:. \label{fixg}
\end{eqnarray}
(\ref{fixg}) states that \epf s behave in the same way as their components, but at the highest (aggregation) level. Important examples include aggregating prices and consumptions, and aggregating wages and labor demands \cite{bkMC,dsMC}. The other assumptions rely on elasticities, substitution elasticities, marginal rates of substitutions, homogeneity and translatability:
\begin{eqnarray}
{{\mathrm{\textsc{e}}}}_{\mu_x}^{x_i} & \defeq & \left(\frac{\mathrm{d} \mu_x}{\mu_x}\right) / \left(\frac{\mathrm{d} x_i}{x_i}\right) \label{defelast}
\end{eqnarray}
is the elasticity of $\mu_x$ with respect to $x_i$,
\begin{eqnarray}
{\mathrm{\textsc{s}}}_{\mu_x}^{x_i \rightarrow x_j} & \defeq & \left( \frac{\partial \mu_x}{\partial x_i} \right) / \left( \frac{\partial \mu_x}{\partial x_j} \right) \label{defmarsub}
\end{eqnarray}
is the marginal rate of substitution of $x_i$ for $x_j$ in $\mu_x$, and
\begin{eqnarray}
{\mathrm{\textsc{e}}}_{\mu_x}^{x_i \rightarrow x_j} & \defeq & \left( \frac{\mathrm{d}(x_j / x_i)}{x_j/x_i} \right) / \left( \frac{\mathrm{d}{\mathrm{\textsc{s}}}_{\mu_x}^{x_i \rightarrow x_j}}{{\mathrm{\textsc{s}}}_{\mu_x}^{x_i \rightarrow x_j}} \right) \label{defelsub}
\end{eqnarray}
is the substitution elasticity of $x_i$ for $x_j$ in $\mu_x$. Finally, function $f(x_1, x_2, ..., x_m)$ is homogeneous of degree $a \in {\mathbb{R}}_*$ if and only if ($\forall \lambda$):
\begin{eqnarray}
f(\lambda x_1, \lambda x_2, ..., \lambda x_m) & = & \lambda^a f(x_1, x_2, ..., x_m) \:\:, \label{defhom}
\end{eqnarray}
and translatable if and only if ($\forall \lambda$):
\begin{eqnarray}
f(\lambda + x_1, \lambda + x_2, ..., \lambda + x_m) & = & \lambda + f(x_1, x_2, ..., x_m) \:\:. \label{deftra}
\end{eqnarray}
\begin{thm}\label{thex}
Let $\mu_x$ be any \lda. The following holds true:
\begin{enumerate}
\item [\bf{(A)}] (\ref{fixg}) holds if and only if $\mu_x$ and $\mu_z$ are CES;
\item [\bf{(B)}] $\sum_{i=1}^{m} {{{\mathrm{\textsc{e}}}}_{\mu_x}^{x_i}} = 1$ if and only if $\mu_x$ is a CES;
\item [\bf{(C)}] $\exists 1\leq i,j\leq m$ such that ${\mathrm{\textsc{e}}}_{\mu_x}^{x_i \rightarrow x_j} = 1$ if and only if $\mu_x$ is a Cobb-Douglas with constant returns to scale;
\item [\bf{(D)}] $\exists 1\leq i,j\leq m$ such that ${\mathrm{\textsc{e}}}_{\mu_x}^{x_i \rightarrow x_j} = a \in {\mathbb{R}}_{+*}$ if and only if $\mu_x$ is a CES;
\item [\bf{(E)}] $\mu_x$ is homogeneous of degree $1$ if and only if it is a CES;
\item [\bf{(F)}] $\mu_x$ is translatable if and only if it is a GEM.
\end{enumerate}
\end{thm}
(proof: see the Appendix). Since Cobb-Douglas with constant returns to scale is a particular case of CES, it meets
simultaneously the assumptions in {\bf A-E}. Theorem \ref{thex} says that if one casts \lda s into an analytical setting compatible with mainstream 
economic assumptions, then the huge set of \lda s reduces to mainstream economic \epf s. This, we think, is a clear-cut position of \lda s
in favor of their economic ``viability''. The rightmost columns of Table \ref{t-epf} summarize the results of Theorem \ref{thex} for
each couple (assumption, \epf), using symbols Y(es), N(o) or L = in the limit.

\subsection{An axiomatization of global transitions costs} 

A natural question is to quantify the global transition
cost ${\mathcal{F}}\left(\bm{x},\bm{y}\right)$ when input $x_i$ shifts to $y_i$ (for $i=1, 2, ..., m$). Assume ${\mathcal{F}}$ is separable, \textit{i.e.} sums input-based contributions:
\begin{eqnarray}
{\mathcal{F}}\left(\bm{x},\bm{y}\right) & \defeq & \sum_{i=1}^{m} {\gamma_i F(x_i, y_i)}\:\:, \label{gtcdef}
\end{eqnarray}
where $F$ quantifies the distortion
of input $i$ during the transition, while $\gamma_i>0$ is its relative \udcdef~(hereafter \udc), and we use
the shorthand vector notation $\bm{x} \defeq (x_1, x_2, ..., x_m)$.
Our goal is to exhibit light conditions on both parameters to fully specify the economic transition, and by the way ground
\lda s as meeting optimality conditions that parallel those of Marshallian and Hicksian demands (\ref{defwalras}, \ref{defhicks}). We start
by $\gamma_i$, and make the assumption that $\gamma_i$ is proportional to $p_i$, 
\textit{i.e.} the \udc~for input $i$ is proportional to its unit price. This is the
\pabdef~setting (\pab).
We make three assumptions on $F$. The two first are structural: \textbf{(i)} it is non negative, \textit{i.e.} it is lower-bounded;
\textbf{(ii)} the local transition cost for input $i$ is zero if and only if
both inputs are the same, \textit{i.e.} if and only if $x_i = y_i$. The third assumption is economic: \textbf{(iii)}
specifies the average input value which minimizes ${\mathcal{F}}$ (\ref{gtcdef}); more precisely, this average
is just the average inputs leveraged by their respective prices. This last assumption connects
input prices to \udc s, and justifies the \pab~setting.

Quite remarkably, these assumptions on \udc s and $F$
are necessary \textit{and} sufficient to completely
shape the setting for economic costs \textit{and} transitions.

\begin{thm}\label{th0}
In the \pab~setting, assume that $F$ is twice continuously differentiable and meets the following assumptions:
\begin{enumerate}
\item [\bf{(i)}] non-negativity: $F(x_i, y_i) \geq 0$; 
\item [\bf{(ii)}] identity of indiscernibles: $F(x_i, y_i) = 0 \Leftrightarrow x_i = y_i$;
\item [\bf{(iii)}] the inputs average minimizes the global transition cost:
\begin{eqnarray}
\arg \min_{y} \frac{1}{\Gamma} \sum_{i=1}^{m} {\gamma_i F(x_i, y)} & = & \frac{1}{P}\sum_{i=1}^{m} {p_i x_i} \label{defmin} \:\:.
\end{eqnarray}
\end{enumerate}
Then $F(x, y) = D_\varphi(x|| y)$ for some strictly convex and differentiable $\varphi$.
\end{thm}
(The proof is a slight variation to that of Theorem 4 in \cite{bgwOT}). Any Bregman divergence satisfies (1.), (2.) and (3.), and so the characterization of $F$ in Theorem \ref{th0} is almost exhaustive given the mild regularity conditions imposed. Let us review the main consequences of Theorem \ref{th0}. Hereafter, we sometimes use the shorthand $w \defeq \sum_{i=1}^{m} {p_i x_i}$.

\subsection{Transition costs top \epfn s and their dual geometries}

Most remarkably, the assumptions of Theorem \ref{th0} yields that the arithmetic average is not the only smallest 
global transition cost: since $D_\varphi$ is not necessarily symmetric, we may also compute the solution to
(\ref{defmin}) in which $x_i$ and $y$ are switched. The solution comes naturally as Legendre duality enters the 
analysis, as we have \cite{nbnOB}:
\begin{eqnarray}
D_\varphi(x_i || y) & = & D_{\varphi^\star}(\textsc{u}(y) || \textsc{u}(x_i))\:\:, \label{ldual}
\end{eqnarray} 
and so the optimum sought immediately follows:
\begin{eqnarray}
\arg \min_{y} \sum_{i=1}^{m} {\gamma_i D_\varphi(y || x_i)} & = & \mu_\varphi \label{defmin2} \:\:.
\end{eqnarray}
This is just the \lda~as formulated in a more general setting in (\ref{bregmeangen}), and justifies
the name ``\lda''. 

From an economic perspective of the \pab~setting, any transition cost grounds an optimal \epf~(the \lda)
which defines lowest cost transitions (its linear expansion path, Lemma \ref{lwh}). The Hicksian demand emerges as a geometric dual (\ref{defmin2})
of the Marshallian demand (\ref{defmin}), a consequence of the property that
transition costs already define dual affine geometries for the inputs, in the
economic input space ($x_i$) and in its image by $\textsc{u}$ (\ref{deru}). This
connection is well studied in differential information geometry \cite{anMO}, and completes
the popular economic duality between both demands \cite{mcwgMT}.

\subsection{Transition costs underlie economic transformation paths}

\begin{landscape}
\begin{figure*}[t]
\centering
\begin{tabular}{cc}
\epsfig{file=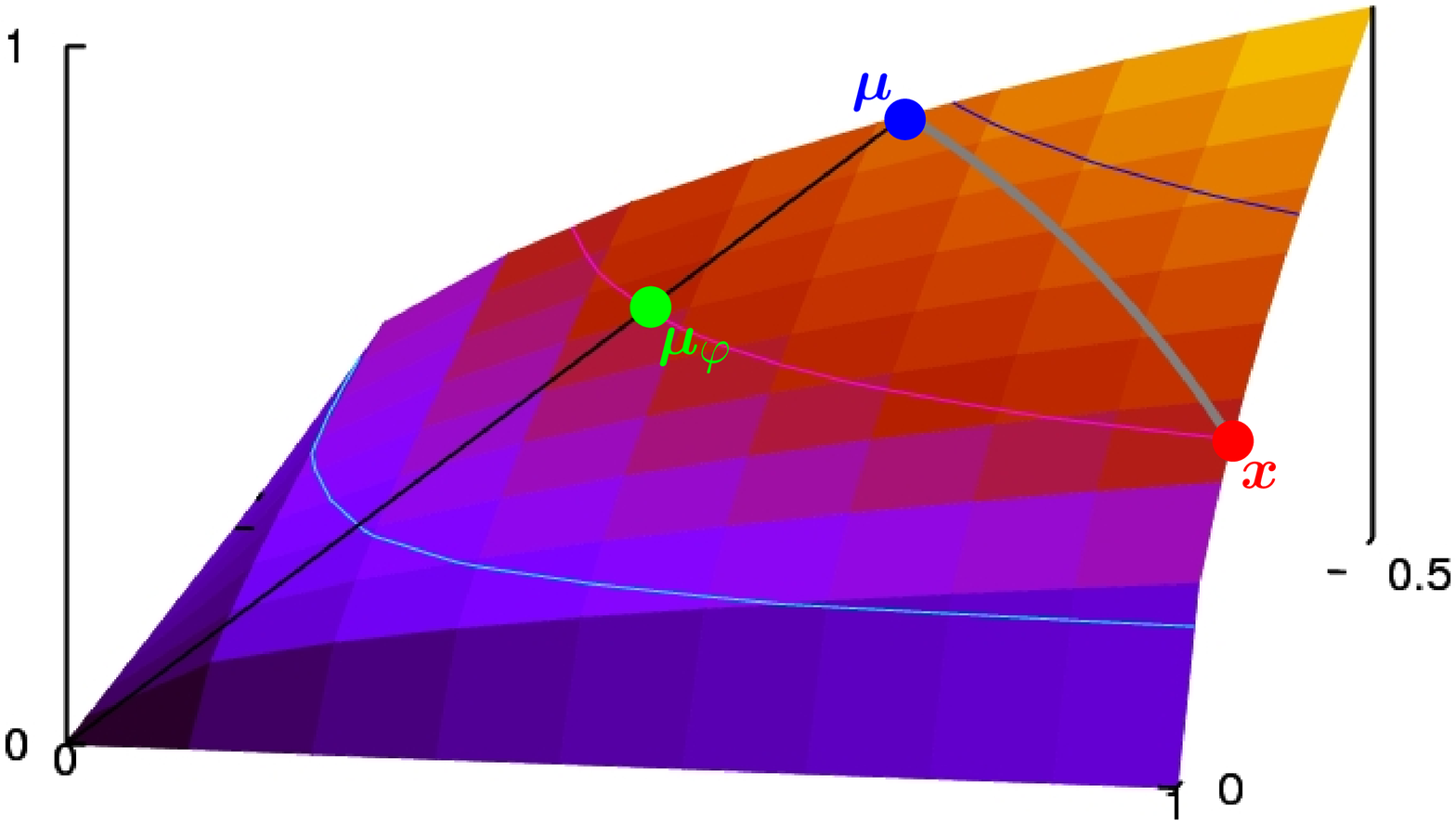,width=8cm} 
& 
\epsfig{file=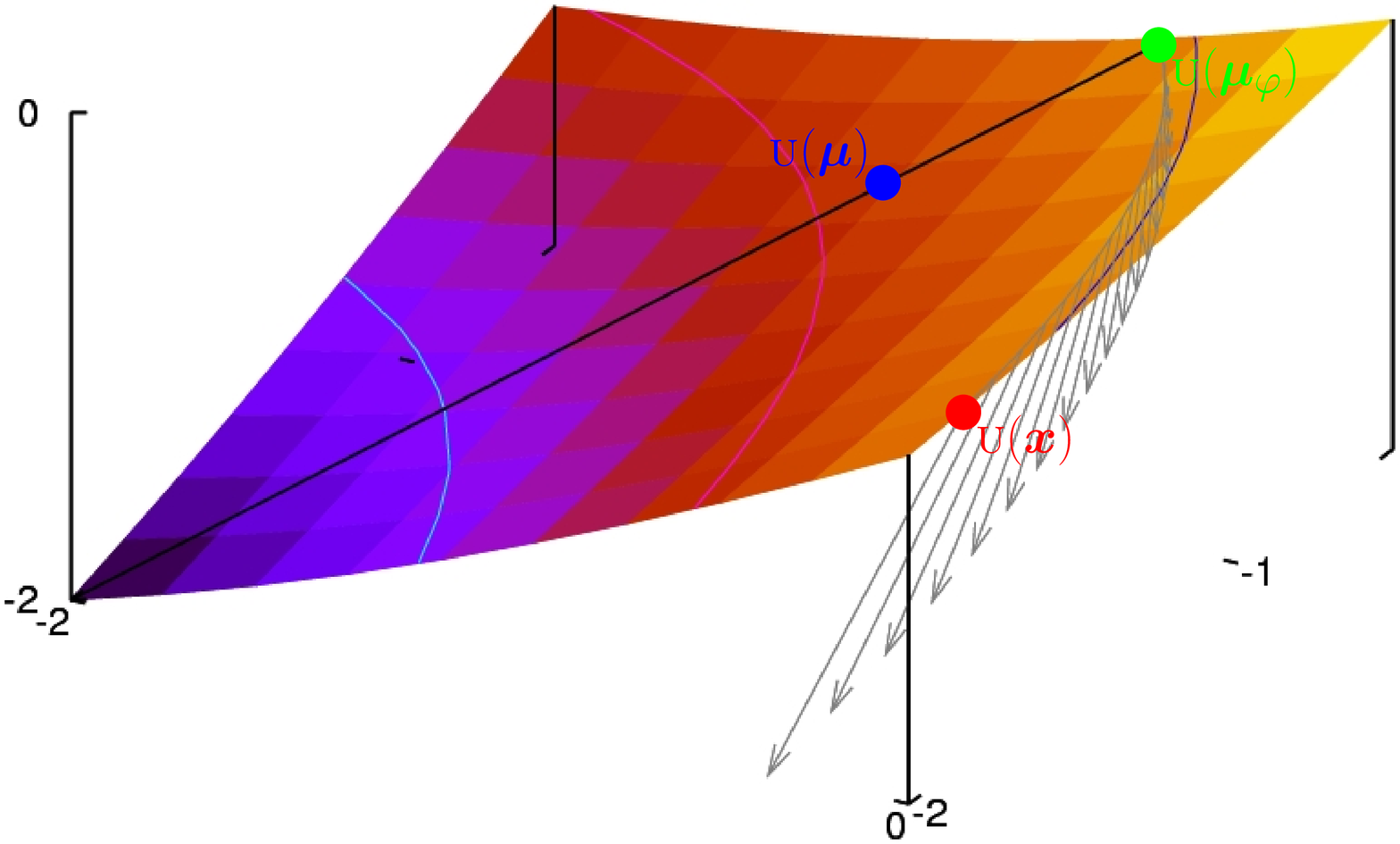,width=8cm}
\end{tabular}
\caption{Left: a Cobb-Douglas ($m=2$); right: its dual (GEM with $\theta = 1$, Table \ref{t-epf}). Isoquants are plotted for both \lda s. On the left, inputs displayed on the \lda~are ${\bm{x}} \defeq (x_1, x_2)$,
${\bm{\mu}}_{\varphi} \defeq (\mu_\varphi, \mu_\varphi)$, ${\bm{\mu}} \defeq (\mu, \mu)$.
The grey paths are the Marshallian (left) and Hicksian (right) paths; the black lines on the \lda s are their expansion paths.\label{f-plotway}}
\end{figure*}
\end{landscape}

We now move onto a less static description of the transition, and show that global transition costs (\ref{gtcdef}) are integrals computed over a particular
\textit{economic transformation path} between $\bm{x}$ and $\bm{y}$. Due to its importance, the Theorem to come
is given in the most general setting: $d$ is arbitrary (Definition \ref{defbreg}), and the divergence is not assumed to be separable \cite{bTR}.
\begin{defn}\label{intloss}
Let $\varphi$ be a function meeting the conditions of Definition \ref{defbreg}. The matching loss parametrized by $\varphi$ is the path integral for vector field $\bm{\upsilon}_\varphi(\bm{z})$:
\begin{eqnarray}
G(\bm{x}_i, \bm{y}_i) & \defeq & \int_{\mathbb{P}} \bm{\upsilon}_\varphi^\top(\bm{z}) \mathrm{d}\bm{z} \:\:, \label{intdist}
\end{eqnarray}
where the vector field and the path are respectively:
\begin{eqnarray}
\bm{\upsilon}_\varphi(\bm{z}) & \defeq & \bm{\nabla}_\varphi(\bm{z}) - \bm{\nabla}_\varphi(\bm{x}_i) \:\:, \label{defupsilon}\\
\mathbb{P} & \defeq & \{\bm{z}(\lambda) \defeq (1-\lambda)\bm{x}_i + \lambda\bm{y}_i, \lambda \in[0,1]\} \:\:. \label{defpath}
\end{eqnarray}
\end{defn}
The link between the path integral and Bregman divergences is stated in the following Theorem.
\begin{thm}\label{lpi}
$G(\bm{x}_i, \bm{y}_i) = D_\varphi(\bm{y}_i || \bm{x}_i)$.
\end{thm}
(proof: see the Appendix). Theorem \ref{lpi} tells us that the economic transition is a linear transformation
from ${\bm{x}}_i$ to ${\bm{y}}_i$. Figure \ref{f-plotway} provides us with an
especially interesting example
of transition path directly mapped on the \epf~in the \pab~setting (left, in grey; the \epf~is a Cobb-Douglas). This path goes to some input 
state ${\bm{x}} \defeq (x_1, x_2)$, starting from its Marshallian
demand on the expansion path (this is point ${\bm{\mu}} \defeq (\mu, \mu)$).
Let us denote this grey path on the \epf~which links ${\bm{\mu}}$ 
to ${\bm{x}}$ the \textit{Marshallian path} of ${\bm{x}}$. As we move along the Marshallian path in the economic input space,
we move along a curve on the dual mean, which goes to $\textsc{u}({\bm{x}})$, starting
from its Hicksian demand  $\textsc{u}(\bm{\mu}_{\varphi})$. We call this dual path, which follows an isoquant of the dual mean, an \textit{Hicksian path} (right picture in
Figure \ref{f-plotway}). We can also define equivalently Hicksian paths in the
economic input space, and Marshallian paths on the dual mean. To make an analogy with physical string deformations, 
the vector field (\ref{defupsilon}) (Figure \ref{f-plotway}, right) is just the force required to keep distorted a string with one endpoint fixed at $\textsc{u}(\bm{\mu}_{\varphi})$,
and the other endpoint somewhere along the Hicksian path. The economic transition cost
is thus analogous to a \textit{work}.

\subsection{Slutsky-type transformations}

\begin{figure}[t]
\centering
\epsfig{file=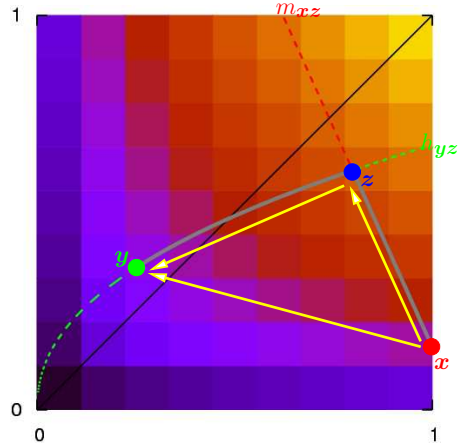,width=6cm} 
\caption{The economic transformation path from $\bm{x}$ to $\bm{y}$ can always be decomposed
in two (\ref{3pts}), involving a particular set of inputs $\bm{z}$, the \bp~of $\bm{x}$ and $\bm{y}$ (yellow arrows; the \epf~is a Cobb-Douglas and $m=2$). Dashed curves $m_{\bm{xz}}$ and $h_{\bm{yz}}$ are orthogonal in that \textit{any} transition from $m_{\bm{xz}}$ to $h_{\bm{yz}}$ admits $\bm{z}$ as \bp~(see text for details).\label{f-slutsky}}
\end{figure}

We show that \textit{any} transformation can be equivalently decomposed in \textit{two} transformations. This decomposition bears surprising similarities with those involved in a fundamental microeconomic equation, Slutsky's identity \cite{mcwgMT}. Our starting point is the following identity, elsewhere known as Bregman triangle equality \cite{nlkMB} (for any $\bm{x}, \bm{y}, \bm{z}$ in $\mathrm{dom}(\varphi)^m$):
\begin{eqnarray}
{\mathcal{F}}(\bm{y}, \bm{x}) & = & {\mathcal{F}}(\bm{y}, \bm{z}) + {\mathcal{F}}(\bm{z}, \bm{x}) + \Delta(\bm{x}, \bm{y}, \bm{z}) \:\:, \label{3pts}
\end{eqnarray}
where $\Delta(\bm{x}, \bm{y}, \bm{z}) \defeq \sum_{i=1}^{m}{\gamma_i (x_i - z_i)(\textsc{u}(z_i) - \textsc{u}(y_i))}$.
For any $\bm{x}$ and $\bm{y}$ in $\mathrm{dom}(\varphi)^m$,
there always exist a $\bm{z}$ in $\mathrm{dom}(\varphi)^m$ for which $\Delta(\bm{x}, \bm{y}, \bm{z}) = 0$ \cite{anMO}. This $\bm{z}$,
 which we call the \bpdef~(\bp) of $\bm{x}$ and $\bm{y}$, yields a decomposition
of ${\mathcal{F}}(\bm{x}, \bm{y})$ in two transitions costs. Figure \ref{f-slutsky} presents an example of this decomposition. 

\begin{figure}[t]
\centering
\epsfig{file=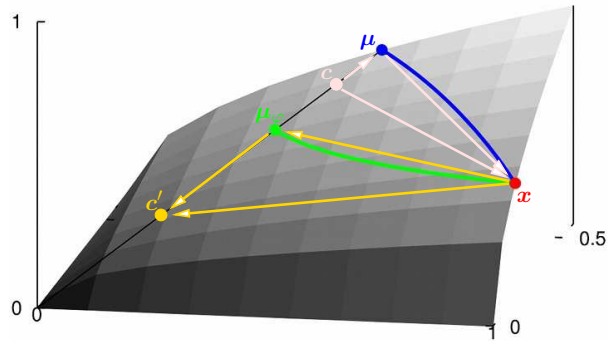,width=8cm} 
\caption{On the Cobb-Douglas \lda, ${\bm{c}} \defeq (c,c)$ and ${\bm{c}}' \defeq (c',c')$ are two points on the expansion path
of the \lda. The transition costs, from ${\bm{c}}'$ to some point ${\bm{x}} \defeq (x_1, x_2)$, or from ${\bm{x}}$ to ${\bm{c}}$,
can be exactly decomposed using only the expansion (black), Marshallian (blue) and Hicksian (green) paths (see text for details, and Figure \ref{f-plotway} for the notations).\label{f-transition}}
\end{figure}

The similarity with the decomposition of Slutsky's identity is striking, yet the framework of (\ref{3pts}) is much different: Slutsky's identity decomposes the variation of the Marshallian demand when input prices change. Hence, the output is completely specified by the change in prices, while (\ref{3pts}) assumes absolutely nothing about the reasons for the output's change. According to Slutsky's identity, the change in the output can be decomposed in a substitution effect between inputs, and an income effect. (\ref{3pts}) tells us similar facts when one endpoint of the transformation is along the expansion path in the \pab~setting (proof immediate from (\ref{ldual}) and (\ref{3pts})).
\begin{thm}\label{decomp}
$\forall c, c' \in {\mathrm{dom}}(\varphi)$,
\begin{eqnarray}
\sum_{i=1}^{m} {\gamma_i D_\varphi (x_i || c)} & = & \sum_{i=1}^{m} {\gamma_i D_\varphi ( x_i || \mu)} + \Gamma D_\varphi (\mu || c)\:\:, \label{rel1}\\
\sum_{i=1}^{m} {\gamma_i D_\varphi (c' || x_i)} & = & \Gamma D_\varphi (c' || \mu_\varphi) + \sum_{i=1}^{m} {\gamma_i D_\varphi (\mu_\varphi || x_i)} \:\:. \label{rel2}
\end{eqnarray}
\end{thm}
Hence, in the \pab~setting, when one endpoint of the transformation lies on the expansion path, the corresponding \bp~is the Marshallian or Hicksian demand, also on the expansion path. Figure \ref{f-transition} presents the two types of transitions, from and to the expansion path: the transformation from ${\bm{x}}$ to ${\bm{c}}'$ on the expansion path (yellow) is the composition of two paths:
\begin{enumerate}
\item [\bf{(a)}] from ${\bm{x}}$ to its Hicksian demand, on the Hicksian path, and
\item [\bf{(b)}] from this Hicksian demand to ${\bm{c}}'$, on the expansion path. 
\end{enumerate}
{\bf (a)} is no more than substitution effect on the transition cost, and {\bf (b)} the income effect on the transition cost. (\ref{rel2}) tells us that leaving the expansion path trades the substitution effect for a \textit{budget} effect in the second stage. 

\subsection{Roy-type elementary variations}

Roy's identity is also fundamental in microeconomics; it says that provided $\mu_\varphi$ meets mild assumptions, we have ${\mathrm{\textsc{s}}}_{\mu_\varphi}^{p_i \rightarrow w} = x_i$ \cite{mcwgMT} (\ref{defmarsub}). It is not hard to prove that \textit{any} elementary input distortion, to or from $\mu_\varphi$, meets Roy's identity when $\mu_\varphi$ does (proofsketch in Appendix).
\begin{thm}\label{iroy} We have ($\forall c \in \mathrm{dom}(\varphi)$):
\begin{eqnarray}
{\mathrm{\textsc{s}}}_{D_\varphi(c || \mu_\varphi)}^{p_i \rightarrow w} & = & {\mathrm{\textsc{s}}}_{D_\varphi(\mu_\varphi || c)}^{p_i \rightarrow w} = {\mathrm{\textsc{s}}}_{\mu_\varphi}^{p_i \rightarrow w} \:\:. \label{eqiroy}
\end{eqnarray}
\end{thm}

\section{Discussion and conclusion}

The anticipations of Charnes \textit{et al.} \cite{ccsAT} are finally not surprising: a significant part of the neoclassical school of economics is about formalizing and aggregating information, and so bonds with information theory had to be expected. What is striking is that those bonds for \epf s come from a weak characterization of --- moreover --- a completely different standpoint on aggregation (dynamic, with geometric flavors). Furthermore, the transition standpoint rejoins quantities and properties popular in the ``static'' (\epf) cases --- Marshallian and Hicksian demands, Slutsky's and Roy's identities, to name a few.

There are various general follow-ups to expect from such a work to continue upon \cite{ccsAT}, two of which appear to be particularly interesting from the economic standpoint. First, an alternative to the transition cost (\ref{gtcdef}) is to compute the maximum cost over inputs. What is economically interesting is that the population minimizer (minmax) is \textit{still} the general \lda~(\ref{bregmeangen}) but on a combinatorial basis of at most $d+1$ inputs \cite{nnFT}: the leveraging coefficients are not the $\gamma_i$s anymore, at most $d+1$ of them are $\neq 0$, they do not have a closed form, but they admit an efficient approximation algorithm \cite{nnFT}. 
Second, alleviating the constraint $d=1$, and even the separability of the global transition cost (\ref{gtcdef}) leads to a rich economic setting of interactions between inputs that deserves further studies.

\bibliographystyle{plain}

\bibliography{../../BIB/bibgen}

\section{Appendix: proofs}

\subsection{Proofsketch of Lemma \ref{lem1}, fourth point}

Without loss of generality, we assume $\Gamma=1$ in (\ref{bregmeangen}). Using the mathematical expectation notation $\es$ in lieu of the average to save space, the concavity of $\bm{\mu}_\varphi$ means $\es_j \bm{\nabla}^{-1}_\varphi(\es_i \bm{\nabla}_\varphi(\bm{x}_{ij}))  \leq \bm{\nabla}_\varphi^{-1}(\es_i \bm{\nabla}_\varphi(\es_j \bm{x}_{ij}))$. Let $\bm{x}_{ij} \defeq \bm{\nabla}^{-1}_\varphi (\bm{x}'_{ij})$ for $\bm{x}'_{ij} \in \mathrm{im}(\bm{\nabla}_\varphi)$. Applying $\bm{\nabla}_\varphi$ on both sides ($\varphi$ is strictly convex, so $\bm{\nabla}_\varphi$ is bijective) and replacing yields:
\begin{eqnarray}
\bm{\nabla}_\varphi(\es_j \bm{\nabla}^{-1}_\varphi(\es_i \bm{x}'_{ij})) & \leq & \es_i \bm{\nabla}_\varphi(\es_j \bm{\nabla}^{-1}_\varphi (\bm{x}'_{ij})) \:\:. \label{defconv}
\end{eqnarray}
Eq. (\ref{defconv}) states the convexity of the \lda~$\tilde{\bm{\mu}} \defeq \bm{\nabla}_\varphi(\es \bm{\nabla}^{-1}_\varphi (\bm{X}))$, but Legendre duality implies $\bm{\nabla}_\varphi = \bm{\nabla}^{-1}_{\varphi^\star}$, and we get $\tilde{\bm{\mu}} = \bm{\mu}_{\varphi^\star}$, the dual of \lda~$\bm{\mu}_\varphi$. The proof starting from the convexity of $\bm{\mu}_\varphi$ follows the same path.

\subsection{Proofsketch of Theorem \ref{thop}}

We only treat the case of MST (point (\textbf{III})). We differentiate the MST in $x_i$. If it is a \lda~with generator $\varphi$, $\mu_x$ must satisfy:
\begin{eqnarray}
\gamma_i \times \frac{-\theta\exp(\theta \gamma_i x_i)}{(1 - \exp(\theta \gamma_i x_i))} \times \mu_x & = & \gamma'_i \times \varphi^{[2]} (x_i) \times \frac{1}{\varphi^{[2]} (\mu_x)}\:\:, \label{condmst}
\end{eqnarray}
with $\gamma'_i$ the \lda~weight for $x_i$. Looking at $\mu_x$, this would imply $\varphi^{[2]} (x) = 1 / x$, from which the simplification of (\ref{condmst}) yields that regardless of the value of $x_i$, the corresponding weights $\gamma_i$ and $\gamma'_i$ must satisfy $- \theta \gamma_i x_i \exp(\theta \gamma_i x_i) = \gamma'_i (1 - \exp(\theta \gamma_i x_i))$, impossible.

\subsection{Proof of Theorem \ref{thex}}

For all six equivalences, implication $\Leftarrow$ is folklore. We prove the reverse implications for points (\textbf{A}), (\textbf{E}) and (\textbf{F}). The remaining
proofs exploit the same tools.\\

(point (\textbf{A})). Since $\mu_x$ is a \lda~with generator $\varphi$, it satisfies:
\begin{eqnarray}
\varphi^{[1]}\left(\mu_x\right) & = & \frac{1}{\Gamma} \sum_{i=1}^{m} {\gamma_i \varphi^{[1]}(x_i)} \label{propc1}\:\:.
\end{eqnarray}
We differentiate (\ref{fixg}) in $x_i$, use (\ref{bregmeangen}), and get:
\begin{eqnarray}
\frac{z_i}{\mu_z} & = & \frac{\gamma_i\varphi^{[2]}(x_i)}{\varphi^{[2]}(\mu_x)}\:\:. \label{genpc}
\end{eqnarray}
We multiply both sides by $x_i$, sum for all $i$, simplify via (\ref{fixg}), rearrange, and get:
\begin{eqnarray}
\mu_x \varphi^{[2]}\left(\mu_x\right) & = & \frac{1}{\Gamma} \sum_{i=1}^{m} {\gamma_i x_i \varphi^{[2]}(x_i)} \:\:. \label{propc2}
\end{eqnarray}
Now, we match (\ref{propc1}) with (\ref{propc2}), and get that $\varphi$ must satisfy:
\begin{eqnarray}
\exists \kappa \in {\mathbb{R}}_* \mbox{ s.t. } \varphi^{[1]}(x) & = & \kappa x \varphi^{[2]}(x), \forall x \in \mathrm{dom}\varphi\:\:. \label{superdom}
\end{eqnarray}
The solution is found to be $\varphi^{[1]}(x) \propto x^\kappa$, i.e.:
\begin{eqnarray}
\varphi(x) & = & \frac{d}{\kappa+1} x^{\kappa+1} \:\:, \label{genphic}
\end{eqnarray} 
with $d \in {\mathbb{R}}_*$ any constant that keeps (\ref{genphic}) convex. Matching (\ref{genphic}) with (\ref{propnces}) implies $\sigma = 1/(1-\kappa)$, and we get the proof that $\mu_x$ is a CES. The other \epf, $\mu_z$, can be found by inspecting (\ref{fixg}) after remarking that partial derivatives on the left and right-hand side must also coincide. After a standard derivation using the general CES form for $\mu_x$ (Table \ref{t-epf}), we obtain that $\mu_z$ is:
\begin{eqnarray}
\mu_z & = & \left( \sum_{i=1}^{m} {\beta_i^\sigma z_i^{1-\sigma}} \right)^{\frac{1}{1-\sigma}} \:\:,\label{defdualp}
\end{eqnarray}
which is also a CES, with generator $\varphi(x) = b x ^{2-\sigma}$, with $b\in {\mathbb{R}}_*$ any constant for which $\varphi$ is convex. 

(point (\textbf{E})). Consider some \lda~$\mu_x$ whose generator is denoted $\varphi$.
Without losing too much generality, we assume that $\varphi$ is twice continuously differentiable, as in Theorem \ref{th0}. (\ref{defhom}) implies:
\begin{eqnarray}
\lefteqn{\left(\varphi^{[1]}\right)^{-1} \left(\frac{1}{\Gamma} \sum_{i=1}^{m} {\gamma_i \varphi^{[1]}(\lambda x_i)}\right)}\nonumber\\
 & = & \lambda^a \left(\varphi^{[1]}\right)^{-1} \left(\frac{1}{\Gamma} \sum_{i=1}^{m} {\gamma_i \varphi^{[1]}(x_i)}\right)\:\:.
\end{eqnarray}
Take some $x_i$, $i=1, 2, ..., m$, and differentiate both sides in $x_i$. We get after simplification:
\begin{eqnarray}
\frac{\lambda \varphi^{[2]}(\lambda x_i)}{\varphi^{[2]}(\lambda^a \mu_x)} & = & \frac{\lambda^a \varphi^{[2]}(x_i)}{\varphi^{[2]}(\mu_x)} \:\:. \label{defc1}
\end{eqnarray}
(\ref{defc1}) implies:
\begin{eqnarray}
\varphi^{[2]} (\lambda x) & = & g(\lambda) \varphi^{[2]}(x)\:\:, \label{propg}
\end{eqnarray} 
for any function $g(\lambda) \in {\mathbb{R}}_*$. Suppose without loss of generality that $g$ is $C_1$, so that we can take the route of the proof of Euler's homogeneous function Theorem. We differentiate (\ref{propg}) in $\lambda$, and take the resulting equation for $\lambda = 1$. We obtain the following PDE:
\begin{eqnarray}
x \varphi^{[3]} (x) - g^{[1]}(1) \varphi^{[2]} (x) & = & 0 \:\:,
\end{eqnarray}
i.e. $\varphi^{[2]}(x) \propto x^\kappa$, where $\kappa \in {\mathbb{R}}_*$ is some constant. We obtain that $\varphi$ is either of the form of (\ref{propnces}), or (\ref{cobb}), the generators of CES or Cobb-Douglas with constant returns to scale (a particular case of CES), as claimed.

(point (\textbf{F})). We take the same route as for the proof of point (\textbf{E}), but differentiating (\ref{deftra}) instead of (\ref{defhom}). (\ref{defc1}) becomes after simplification:
\begin{eqnarray}
\frac{\varphi^{[2]}(\lambda + x_i)}{\varphi^{[2]}(x_i)} & = & \frac{\varphi^{[2]}(\lambda + \mu_x)}{\varphi^{[2]}(\mu_x)} \:\:, \label{defc2}
\end{eqnarray}
which yields $\varphi^{[2]}(x + \lambda) = g(\lambda) \varphi^{[2]}(x)$ for some function $g$, and, taking $x=0$, brings $g(\lambda) = \varphi^{[2]}(\lambda) / \varphi^{[2]}(0)$. We obtain $\varphi^{[2]}(x + \lambda) = \varphi^{[2]}(\lambda) \varphi^{[2]}(x) / \varphi^{[2]}(0)$, implying $\varphi^{[2]}(x) \neq 0$, and thus $\varphi^{[2]}(2x) = (\varphi^{[2]})^2(x) / \varphi^{[2]}(0)$. The change of variable $g(x) \defeq \ln(|\varphi^{[2]}(x)|)$ yields:
\begin{eqnarray}
g(2x) = 2g(x) - g(0)\:\:,
\end{eqnarray}
and thus $g^{[1]}(2x) = g^{[1]}(x) = K$, a constant. Taking the route back to $\varphi^{[1]}$, we easily obtain:
\begin{eqnarray*}
\varphi^{[1]}(x) & = & a \exp(\theta x + b) + c\:\:, a \in {\mathbb{R}}_{+*}, \theta \in {\mathbb{R}}_*, b, c \in {\mathbb{R}}\:\:,
\end{eqnarray*}
from which we recover the \lda~of the Generalized Exponential Mean (Table \ref{t-epf}).

\subsection{Proof of Theorem \ref{lpi}}

One easily recovers Bregman divergences as $\mathrm{d}\bm{z} = (\bm{y}_i - \bm{x}_i)\mathrm{d}\lambda$, and the integral becomes:
\begin{eqnarray}
G(\bm{x}_i, \bm{y}_i) & = & \int_{0}^{1} {(\bm{y}_i - \bm{x}_i)^\top (\bm{\nabla}_\varphi(\bm{z}(\lambda)) - \bm{\nabla}_\varphi(\bm{x}_i))\mathrm{d}\lambda} \label{scal}\\
 & = & - (\bm{y}_i - \bm{x}_i)^\top \bm{\nabla}_\varphi(\bm{x}_i) \nonumber\\
 & & + \int_{0}^{1} {(\bm{y}_i - \bm{x}_i)^\top \bm{\nabla}_\varphi((1-\lambda)\bm{x}_i + \lambda\bm{y}_i) \mathrm{d}\lambda} \nonumber\\
 & = & - (\bm{y}_i - \bm{x}_i)^\top \bm{\nabla}_\varphi(\bm{x}_i) + \int_{0}^{1} { \mathrm{d} \varphi((1-\lambda)\bm{x}_i + \lambda\bm{y}_i)}   \nonumber\\
 & = & - (\bm{y}_i - \bm{x}_i)^\top \bm{\nabla}_\varphi(\bm{x}_i) + \varphi(\bm{y}_i) - \varphi(\bm{x}_i) \nonumber\\
 & = & D_\varphi(\bm{y}_i || \bm{x}_i)\:\:. \label{pfint}
\end{eqnarray}

\subsection{Proofsketch of Theorem \ref{iroy}}

The proof is immediate once we remark that the derivative of the distortion is proportional to the derivative of the \lda:
\begin{eqnarray}
\frac{\partial D_\varphi(c || \mu_\varphi)}{\partial u} & = & - \varphi^{[2]}(\mu_\varphi) (c - \mu_\varphi) \frac{\partial \mu_\varphi}{\partial u}\:\:, \label{p1}\\
\frac{\partial D_\varphi(\mu_\varphi || c)}{\partial u} & = &  (\varphi^{[1]}(\mu_\varphi) - \varphi^{[1]}(c)) \frac{\partial \mu_\varphi}{\partial u}\:\:. \label{p2}
\end{eqnarray}

\begin{landscape}
\begin{table*}
\caption{Some popular Bregman divergences $D_\varphi(x_i || y_i)$.\label{t-exe}}
\begin{tabular}{llll}\hline \hline
$\mathrm{dom}(\varphi)$ & $\varphi(x)$  & $D_\varphi(x_i|| y_i)$ & Divergence name\\ \hline\hline
${\mathbb{R}}$ & $x^2$ & $(x_i - y_i)^2$ & Squared Euclidean norm \\ \hline
${\mathbb{R}}_+$ & $x \log x - x$ & $x_i \log\frac{x_i}{y_i} - x_i + y_i$ & Kullback-Leibler divergence \\ \hline
${\mathbb{R}}_{+*}$ & $-\log x$ & $\frac{x_i}{y_i} - \log\frac{x_i}{y_i} - 1$ & Itakura-Saito divergence \\ \hline
${\mathbb{R}}$, $\alpha\in (-1,1)$ & $\frac{4}{1-\alpha^2}\left(x-x^{\frac{1+\alpha}{2}}\right)$ & $\frac{2}{1+\alpha}x_i^{\frac{1+\alpha}{2}} + \frac{2}{1-\alpha}y_i x_i^{\frac{\alpha-1}{2}} - \frac{4}{1-\alpha^2}y_i^{\frac{1+\alpha}{2}}$ & Amari $\alpha$-divergence \\ 
${\mathbb{R}}_{+*}$, $\alpha\rightarrow-1$ & \multicolumn{2}{c}{$=$Kullback-Leibler divergence$(x_i || y_i)$} &  \\
${\mathbb{R}}_{+}$, $\alpha\rightarrow1$ & \multicolumn{2}{c}{$=$Kullback-Leibler divergence$(y_i || x_i)$} & \\ \hline
${\mathbb{R}}$, $\alpha\in (0,1)$ & $\frac{-x^\alpha +\alpha x - \alpha + 1}{\alpha(1-\alpha)}$ & $\frac{1}{\alpha(1-\alpha)}(x_i^\alpha - y_i^\alpha -\alpha y_i x_i -\alpha x^2_i)$ & Bregman-Csisz\'ar divergence \\ 
${\mathbb{R}}_{+*}$, $\alpha\rightarrow0$ & \multicolumn{2}{c}{$=$Itakura-Saito divergence$(x_i || y_i)$} &  \\
${\mathbb{R}}_{+}$, $\alpha\rightarrow1$ & \multicolumn{2}{c}{$=$Kullback-Leibler divergence$(x_i || y_i)$} &  \\ \hline \hline
${\mathbb{R}}$, $\alpha\in (0,1)$ & $\frac{(x^{\frac{1}{\alpha}} + 1)^\alpha - 2^\alpha}{2(1-\alpha)}$ & $\frac{1}{2(1-\alpha)}( (x_i^{\frac{1}{\alpha}} + 1)^\alpha - (y_i^{\frac{1}{\alpha}} + 1)^{\alpha-1}( 1 - x_i y_i^{\frac{1}{\alpha}-1} + 2y_i^{\frac{1}{\alpha}}) )$ & Arimoto divergence \cite{lvCS} \\ 
${\mathbb{R}}$, $\alpha\rightarrow0$ & \multicolumn{2}{c}{$=F_1$ divergence$(x_i || y_i)$} &  \\
${\mathbb{R}}_{+*}$, $\alpha\rightarrow1$ & \multicolumn{2}{c}{$=$Bayesian divergence$(x_i || y_i)$} & \\ \hline \hline
\end{tabular}
\end{table*}

\begin{table*}
\caption{Famous economic production functions ($\beta_i, \beta_{ij} > 0, \forall i, j$), along with a summary of our results in Theorems \ref{thop} and \ref{thex} (see text for details). MSTs are given up to some eventual variable change.\label{t-epf}}
\begin{tabular}{ll|c|cccccc}\hline \hline
\epf~$\mu_x$ & Name & Optimality & \multicolumn{6}{|c}{Exhaustivity (Th. \ref{thex})}\\ 
    &      & (Th. \ref{thop}) & (\textbf{A}) & (\textbf{B}) & (\textbf{C}) & (\textbf{D}) & (\textbf{E}) & (\textbf{F})\\  \hline
$\left(\sum_{i=1}^{m} \beta_i x_i^\frac{\sigma-1}{\sigma} \right)^{\frac{\sigma}{\sigma - 1}}$ & Constant Elasticity of Substitution & Y & Y & Y & N & Y & Y & N\\
 & (CES, $\sigma \in {\mathbb{R}}_* \backslash \{1\}$) \cite{acmsCL} & & & & & & \\ \hline
$\prod_{i=1}^{m} {x_i^{\beta_i}}$  & Cobb-Douglas with constant & Y & Y & Y & Y & Y & Y & N \\
& returns to scale ($\sum_i {\beta_i} = 1$) \cite{cdAT,mcwgMT} & & & & & & & \\ \hline
$(1/\theta) \log\left(\sum_{i=1}^{m} {\beta_i \exp(\theta x_i)}\right)$ & Generalized Exponential Mean & Y & N & N & N & N & N & Y\\
 & (GEM, $\theta \in {\mathbb{R}}_*$) \cite{bbdII} & & & & & & \\ \hline
$\min_i \{\beta_i x_i\}$ & Leontief \cite{lTS} & L & L & L & N & L & L & N \\ \hline
$\exp(\beta_0 + \sum_{i=1}^{m}{\beta_i\log{x_i}}$ & Translog \cite{cjlCD}& N & N & N & N & N & N & N\\
$+ \sum_{i=1}^{m}{\sum_{j=1}^{m} {\beta_{ij} \log x_i \log x_j}})$ & & & & & & &\\ \hline
$\prod_{i=1}^{m} {(1 - \exp(\theta \beta_i x_i))}$ & Mitscherlich-Spillman-von Th{\"u}nen  & N & N & N & N & N & N & N\\ 
 &  (MST, $\theta \in \{-1,+1\}$) \cite{mDG,sTL,tDI} & & & & & & & \\ \hline \hline
\end{tabular}
\end{table*}
\end{landscape}

\end{document}